\documentclass[journal]{IEEEtran}
\usepackage{ifpdf}
\usepackage{url}
\ifpdf
\else
\fi
\usepackage{acro}
\usepackage{cite}
\usepackage{balance}
\usepackage{bm,comment,color}

\ifCLASSINFOpdf
  \usepackage[pdftex]{graphicx}
  \graphicspath{{../pdf/}{../jpeg/}}
  \DeclareGraphicsExtensions{.pdf,.jpeg,.png}
\else
  \usepackage[dvips]{graphicx}
  \graphicspath{{../eps/}}
  \DeclareGraphicsExtensions{.eps}
\fi
\usepackage{amsmath}
\usepackage{algpseudocode}% http://ctan.org/pkg/algorithmicx
\algtext*{EndWhile}% Remove "end while" text
\algtext*{EndIf}% Remove "end if" text
\algtext*{EndFor}
\usepackage[ruled,vlined]{algorithm2e}

\usepackage{array}
\usepackage{amsfonts} 
\usepackage{amssymb}
\usepackage{esint} % various fancy integral symbols
\usepackage{units}
\usepackage{multirow}
\usepackage{comment}

\usepackage{bbm}

\usepackage{color}
\usepackage{standalone}

\usepackage{pgfplots}
\usepackage{tikz}
\usetikzlibrary{calc}
\makeatletter
\newcommand{\gettikzxy}[3]{%
  \tikz@scan@one@point\pgfutil@firstofone#1\relax
  \edef#2{\the\pgf@x}%
  \edef#3{\the\pgf@y}%
}
\usetikzlibrary{spy,backgrounds}
\pgfplotsset{compat=newest}
\usetikzlibrary{plotmarks}
\usetikzlibrary{arrows.meta}
\usepgfplotslibrary{patchplots}
\usepackage{grffile}
\newlength\fheight 
\newlength\fwidth 
\usepgfplotslibrary{fillbetween}

\usepackage{tabu,longtable}

\ifCLASSOPTIONcompsoc
 \usepackage[caption=false,font=normalsize,labelfont=sf,textfont=sf]{subfig}
\else
 \usepackage[caption=false,font=footnotesize]{subfig}
\fi
\usepackage{stfloats}
\usepackage[hidelinks]{hyperref}
\usepackage{xcolor}
\long\def\comment#1{}

\newfont{\bbb}{msbm10 scaled 700}

\newfont{\bb}{msbm10 scaled 1100}

\renewcommand{\arg}{{\hbox{arg}}}

\usepackage{graphicx}      % For including images
\usepackage{booktabs}

\usepackage{amsthm}

\usepackage{caption}
\usepackage{subcaption}
\begin{document} 
\bstctlcite{IEEEexample:BSTcontrol}

\title{RA-LWLM: Retrieval-Augmented In-Context Localization with Wireless Foundation Models}

\author{
Guangjin Pan, \IEEEmembership{Member, IEEE}, Hui Chen, \IEEEmembership{Member, IEEE},  Hei Victor Cheng, \IEEEmembership{Member, IEEE}, \\
Henk Wymeersch, \IEEEmembership{Fellow, IEEE}
\thanks{G. Pan, H. Chen, and H. Wymeersch are with the Department of Electrical Engineering, Chalmers University of Technology, 41296 Gothenburg, Sweden (email: {guangjin.pan; hui.chen; henkw}@chalmers.se).}
\thanks{H. V. Cheng is with the Department of Electrical and Computer
Engineering, Aarhus University, Denmark. (email: hvc@ece.au.dk).}

\thanks{This work was supported in part by a grant from the Chalmers AI Research
 Center Consortium (CHAIR), by the SNS JU project 6G-DISAC under the EU's Horizon Europe research and innovation Program under Grant Agreement No. 101139130, the Swedish Foundation for Strategic Research (SSF) (grant FUS21-0004, SAICOM),  and Chalmers Areas of Advance in ICT and Transport. The computations were enabled by resources provided by the National Academic Infrastructure for Supercomputing in Sweden (NAISS), partially funded by the Swedish Research Council through grant agreement No. 2022-06725. (Corresponding author: Hui Chen).
}
}

\maketitle

\begin{abstract}
Wireless localization is a fundamental capability of sixth-generation (6G) networks. Conventional model-based methods require accurate modeling of the propagation environment and degrade in complex multipath and non-line-of-sight scenarios, while learning-based methods couple model parameters tightly to the training scene, requiring costly retraining whenever the base station (BS) configuration or propagation environment changes. In this paper, we propose RA-LWLM, a retrieval-augmented in-context localization framework that achieves training-free cross-scene adaptation by externalizing scene-specific information into a per-scene fingerprint database rather than encoding it in model weights. The framework consists of three components: a frozen wireless foundation model (FM) encoder that maps raw channel state information into a scene-agnostic representation; a retrieval module that selects the most informative references from the per-scene database via similarity search in the representation space; and a transformer-based in-context learning (ICL) module that fuses the query with the retrieved references to predict the user equipment (UE) position. To accommodate varying retrieval quality and propagation complexity across queries, the ICL module adopts a mixture-of-experts design in which experts specialize in different context sizes and are softly combined by a learnable selector. Extensive ray-tracing-based experiments across heterogeneous scenes with diverse BS configurations show that RA-LWLM achieves nearly identical accuracy on seen and unseen scenes without any per-scene retraining, substantially outperforming end-to-end and FM-based baselines. These results validate the proposed retrieval-augmented in-context paradigm as a scalable solution for cross-scene localization in 6G networks.
\end{abstract}

\IEEEpeerreviewmaketitle
\acresetall 

\section{Introduction}
Accurate user equipment (UE) localization is a fundamental capability of sixth-generation (6G) wireless networks, underpinning emerging applications such as autonomous systems, extended reality, the low-altitude economy, and location-aware network optimization \cite{talvitie2023orientation, pan2025ai,yang2024positioning,jiao2026opennavmap}. In addition to supporting these applications, accurate position information enhances system-level functionalities of the 6G networks, including mobility management~\cite{di2014location}, interference mitigation~\cite{chen2024location}, and resource scheduling~\cite{kwon2023integrated}. Therefore, achieving accurate UE localization that generalizes across diverse deployment scenarios is one of the key enablers of the 6G vision \cite{pan2025ai}.

To meet these requirements, two main paradigms have been investigated. Model-based localization methods exploit geometric relationships such as time-of-arrival (ToA), time-difference-of-arrival (TDoA), and angle-of-arrival (AoA) to triangulate the UE position, and can achieve high accuracy when the propagation environment is well characterized \cite{yang2024positioning}. However, these methods require accurate modeling of the propagation environment and tend to degrade in multipath-rich and non-line-of-sight (NLOS) scenarios \cite{chen2022tutorial,Wang2024Multi}. In contrast, learning-based localization delegates the complex channel-to-position mapping to data-driven models, allowing it to fully exploit the rich features embedded in channel state information (CSI) \cite{pan2025ai,Castillo2025MIMO}. 

\subsection{Related Work}

Building on the learning-based paradigm, a large body of work has investigated how to map CSI measurements to UE positions through deep neural networks. Early studies formulate localization as a regression or classification task on hand-crafted signal features, such as received signal strength and CSI statistics, and feed them into multi-layer perceptrons (MLPs) to predict UE coordinates \cite{zhang2016deep}. To better capture the spatial-frequency structure of CSI, subsequent works directly take the raw CSI matrix as input and design convolutional neural network (CNN)-based fingerprinting models that learn discriminative features in an end-to-end manner, such as \cite{wu2021learning,pan2020High}. More advanced neural architectures, including graph neural networks \cite{wang2024graph,Yan2025GNN}, transformers \cite{xu2024swin}, and mixture-of-experts (MoE) \cite{wang2025spatial}, have been further introduced to strengthen the channel-to-position mapping and achieve higher localization accuracy. Nevertheless, learning-based methods rely on offline fingerprint collection with ground-truth labels and model training, and the resulting model parameters are tightly coupled to the base station (BS) configuration (e.g., antenna number, bandwidth, orientation) and the propagation environment of the training scene \cite{pan2025ai}. Consequently, the same set of model parameters generalizes poorly across scenes, and each new deployment requires re-collecting fingerprints and retraining the model, which severely limits the scalability of learning-based localization in large-scale heterogeneous 6G networks.

To alleviate this scalability bottleneck, researchers have proposed several distinct approaches. One direction is channel charting \cite{studer2018channel,stephan2024angle,ferrand2021triplet,zhang2025unilocpro,aly2023model,mateos2025positioning}, which constructs a low-dimensional pseudo-position representation from CSI through dimensionality reduction techniques, eliminating the need for labeled fingerprints. The seminal work in \cite{studer2018channel} proposes to learn a chart by preserving CSI-domain dissimilarities, so that geometrically nearby UEs are mapped to nearby chart points. Building on this idea, subsequent works improve the chart quality by introducing more informative dissimilarity metrics, e.g., angle-delay profile-based and timestamp-aided metrics \cite{stephan2024angle}, or by adopting more expressive neural backbones such as Siamese and triplet networks \cite{ferrand2021triplet}. More recent efforts further unify channel charting with model-based geometric priors to improve absolute positioning accuracy. However, channel charting fundamentally produces only a relative representation, and the quality of this representation hinges on the choice of dissimilarity metric used during training \cite{zhang2025unilocpro}. Designing a dissimilarity metric that is both physically meaningful and well-suited for localization remains a significant open challenge. Moreover, deploying channel charting to a new scene still requires retraining a scene-specific charting network on newly collected unlabeled CSI and a calibration step with anchor points to recover absolute UE coordinates.

Another direction to reduce the labeled-data cost of adapting to a new scene is to transfer knowledge from previously seen scenes based on transfer learning or meta learning \cite{Foliadis2025Transfer,Cui2022transfer,gao2023metaloc,yan2025attentional}. Transfer learning approaches typically pretrain a fingerprinting model on one or several source scenes and then fine-tune part of the network on a small set of labeled samples from the target scene \cite{Foliadis2025Transfer,Cui2022transfer}. Meta-learning methods learn an initialization or adaptation strategy from a distribution of training scenes, so that the model can quickly adapt to unseen scenes with only a few gradient steps \cite{gao2023metaloc,yan2025attentional}. Although these methods substantially reduce the amount of target-scene data required, deploying them to a new scene still entails a fine-tuning step, and both labeled samples and computational resources have to be expended for every new scene.

More recently, inspired by the success of foundation models (FMs) in language and vision, a number of wireless FMs \cite{jiang2025towards,guo2026large,liu2026self,Salihu2024Self-Supervised,pan2025large} have been proposed to learn universal channel representations through self-supervised pretraining on large-scale unlabeled CSI. Unlike transfer learning, which produces a localization model for fine-tuning, FM-based pretraining is task-agnostic and produces general-purpose channel features, and has been shown to deliver better downstream performance~\cite{liu2026self,Salihu2024Self-Supervised,pan2025large}. For example, \cite{liu2026self} pretrains a transformer-based encoder via masked CSI reconstruction across diverse scenes, while \cite{Salihu2024Self-Supervised} learns transferable features through contrastive objectives over different views of the same channel. The large wireless localization model (LWLM)~\cite{pan2025large} further introduces a hybrid pretraining method that combines generation-based reconstruction and contrastive-based learning to obtain an FM tailored to wireless localization. Once pretrained, the FM encoder can be reused as a generic feature extractor for various downstream tasks, including channel prediction, beam management, and localization \cite{jiang2025towards,guo2026large}. Although FMs can extract more general and transferable channel features, when applied to localization, existing FM-based end-to-end localization methods still need to train a scene-specific decoder on labeled fingerprints, since the decoder must learn the mapping from FM features to absolute coordinates within the geometry of each particular scene. Consequently, the cross-scene adaptation problem is not fundamentally resolved. This motivates rethinking the way scene-specific information is incorporated into the localization framework.

The above limitations motivate a fundamentally different design philosophy: instead of encoding scene-specific information into model weights, can we externalize it into a per-scene reference database that is queried at inference time, so that adapting to a new scene reduces to swapping the database rather than retraining the model? This idea is inspired by two recent paradigms that have transformed natural language processing. Retrieval-augmented generation (RAG) grounds the predictions of large language models (LLMs) on external knowledge bases retrieved at inference time, allowing the model to incorporate up-to-date or domain-specific information without parameter updates \cite{yu2024rankrag,lewis2020retrieval}. On the other hand, In-context learning (ICL) \cite{yang2024context} makes predictions on a query sample by learning the input-output relationship from a few reference examples provided in the prompt, again without any parameter updates. Together, RAG and ICL offer a compelling blueprint for training-free adaptation, in which task- or domain-specific knowledge is supplied externally rather than baked into the model \cite{ram2023context}. While retrieval-based ideas have recently started to be explored in wireless communications for tasks such as resource allocation~\cite{zeeshan2025llm} and semantic communication~\cite{tang2025retrieval}, their systematic application to wireless localization has not yet been fully investigated. A recent attempt~\cite{huang2026retrieval} proposes a channel-charting-based graph neural network framework, in which channel charting is used for dimensionality reduction and retrieval, while the GNN aggregates information from the retrieved reference fingerprints. Although this approach improves localization accuracy over end-to-end learning, the cross-scene generalization problem is not systematically addressed and discussed.

\subsection{Contributions}

Inspired by these observations, in this paper, we propose a retrieval-augmented LWLM (RA-LWLM) that combines a wireless FM with the RAG and ICL paradigms to achieve training-free cross-scene localization. The key insight is to decouple the scene-invariant channel-to-feature mapping from the scene-specific feature-to-position mapping: the former is learned once by a pretrained FM and shared across all scenes, while the latter is supplied at inference time by retrieving labeled references from a per-scene database. Therefore, adapting to a new scene reduces to refreshing the database, with no parameter update required. The main contributions of this paper are summarized as follows.

\begin{itemize}
    \item \textbf{Retrieval-augmented localization framework.} We recast the fingerprint wireless localization task as a retrieval-augmented inference problem in which environment-specific information is externalized into a per-scene fingerprint database rather than encoded into model weights. Specifically, for each query channel, we retrieve highly similar reference samples from the channel–position database and design an ICL network that performs position estimation conditioned on these references. Within this framework, scene heterogeneity, including diverse BS configurations and propagation environments, is handled uniformly under a single training-free adaptation scheme, eliminating the per-scene retraining required by conventional fingerprinting methods.
    
    \item \textbf{Foundation-model-based representation for retrieval.} We leverage the pretrained FM encoder to map raw CSI into a scene-agnostic representation that supports efficient similarity search on a fixed-dimensional vector regardless of the underlying CSI dimension. Performing retrieval in this learned representation space resolves the high-dimensionality and cross-scene comparability issues that hinder direct retrieval on raw CSI.
    
    \item \textbf{A mixture-of-experts (MoE)-based  ICL localization module with adaptive context size.} We design a transformer-based ICL module whose experts specialize in different context sizes and are softly combined by a learnable selector. With weighted-centroid centering and spatial normalization, each expert predicts only a small position residual from its retrieved references, encouraging transferable spatial reasoning. A two-stage training scheme, consisting of per-expert pretraining followed by selector-only routing training, is further proposed to ensure stable training of both the ICL experts and the routing module.

    \item \textbf{Comprehensive evaluation on heterogeneous scenes.} We conduct extensive ray-tracing-based experiments across diverse BS configurations and propagation environments. The results show that RA-LWLM achieves nearly identical accuracy on seen and unseen scenes, substantially outperforming both conventional fingerprinting and FM-based baselines. Beyond performance improvements, our evaluation reveals two distinct roles of training resources, where the number of training scenes primarily improves cross-scene generalization while the per-scene database size primarily reduces the absolute localization error within any given scene. This separation phenomenon provides a practical guideline for allocating the data collection budget in real deployments.

\end{itemize}

The remainder of this paper is organized as follows. Section II presents the system model and formulates the retrieval-augmented localization problem. Section III details the proposed RA-LWLM framework, including the FM-based representation extraction, the retrieval module, and the ICL localization module. Section IV describes the training procedure and model architecture. Section V presents the evaluation results, followed by conclusions in Section VI.

%====================================================================
\section{System Model and Problem Formulation}
\label{sec:SystemModel}
%====================================================================

In practice, different BS configurations and propagation environments across deployment scenes lead to heterogeneous channel characteristics, which pose a significant generalization challenge for data-driven localization. We consider $S$ deployment scenes, collectively denoted by $\mathcal{S} = \{1, 2, \dots, S\}$. To establish a consistent parameterization across scenes, since the BS deployment is typically known, without loss of generality, we translate the horizontal 2D coordinate of the BS to the origin in each scene $s \in \mathcal{S}$, so that the position of BS $s$ is $\bm{p}_s^{\text{bs}} = [0, 0, z_s^{\text{bs}}]^\top$, where $z_s^{\text{bs}}$ denotes the scene-dependent BS antenna height. In this work, we estimate only the horizontal UE position; the horizontal UE coordinate within scene $s$ is denoted by $\bm{p}_s^{\text{ue}} = [x_s^{\text{ue}}, y_s^{\text{ue}}]^\top \in \mathbb{R}^2$, which is the unknown to be estimated.\footnote{Although we focus on 2D localization throughout this work, the proposed framework can be extended to 3D localization by including the UE altitude.}

On top of this spatial setup, each scene $s$ adopts an uplink multiple-input multiple-output orthogonal frequency-division multiplexing (MIMO-OFDM) system in which a single-antenna UE communicates with a BS equipped with a uniform linear array (ULA). The number of BS antennas $N^{\text{ant}}_s$ and the total system bandwidth $B^{\text{bw}}_s$ are both treated as configurable parameters that may differ across scenes. The bandwidth is divided into $N^{\text{subc}}_s$ orthogonal subcarriers with spacing $\Delta^{f}_s = B^{\text{bw}}_s / N^{\text{subc}}_s$. In addition to these BS-side configurations, each scene is further characterized by a distinct propagation environment $\mathcal{E}_s$ and a BS orientation angle $\varphi^{\text{az}}_s$, both of which contribute to the diversity of channel characteristics across scenes.

\subsection{Channel Model}
\label{subsec:ChannelModel}

In scene $s$, the BS observes a CSI matrix $\bm{H}_s = [\bm{h}_{s,1}, \dots, \bm{h}_{s,N^{\text{subc}}_s}] \in \mathbb{C}^{N^{\text{ant}}_s \times N^{\text{subc}}_s}$, whose $m$-th column $\bm{h}_{s,m} \in \mathbb{C}^{N^{\text{ant}}_s}$ corresponds to the channel frequency response (CFR) at the $m$-th subcarrier. Following the standard multipath model, the CFR vector is expressed as
\begin{align}
    \bm{h}_{s,m} = \sum_{l=1}^{L} \alpha_{l} \, \bm{a}_s^{\text{bs}}(\theta_{l}) \, e^{-j 2\pi m\Delta^{f}_s \tau_{l}} + \bm{n}_{s,m},
    \label{eq:channel}
\end{align}
where $L$ denotes the number of multipath components (MPCs), and $\alpha_{l} \in \mathbb{C}$, $\tau_{l}$, and $\theta_{l}$ represent the complex gain, propagation delay, and AoA of the $l$-th MPC, respectively. For ease of understanding, we simplify the notation of these multipath parameters. In reality, $L$, $\alpha_{l}$, $\tau_{l}$, and $\theta_{l}$ all depend on the BS configuration (including the BS orientation $\varphi^{\text{az}}_s$), the UE position $\bm{p}_s^{\text{ue}}$, and the propagation environment $\mathcal{E}_s$. The term $\bm{n}_{s,m} \in \mathbb{C}^{N^{\text{ant}}_s}$ denotes the additive white Gaussian noise (AWGN) vector. The ULA steering vector $\bm{a}_s^{\text{bs}}(\cdot) \in \mathbb{C}^{N^{\text{ant}}_s}$ is defined in the local coordinate system of the BS array and takes the standard form
$\bm{a}_s^{\text{bs}}(\theta) = \big[1,\, e^{-j \frac{2\pi d}{\lambda}\sin\theta},\, \dots,\, e^{-j \frac{2\pi d}{\lambda}(N^{\text{ant}}_s - 1)\sin\theta}\big]^\top$,
where $\theta$ denotes the AoA, $\lambda$ is the carrier wavelength, and $d$ is the antenna spacing.

\subsection{Dataset Model}
\label{subsec:Dataset}

For each scene $s$, we assume that the CSI measurements and the corresponding UE positions are jointly drawn from a scene-specific distribution
\begin{align}
    p_s(\bm{H},\, \bm{p}^{\text{ue}}) \,\triangleq\, p\!\left(\bm{H},\, \bm{p}^{\text{ue}} \,\big|\, \bm{c}_s,\, \mathcal{E}_s\right),
    \label{eq:scene_dist}
\end{align}
where $\bm{c}_s = \{z_s^{\text{bs}},\, \varphi^{\text{az}}_s,\, N^{\text{ant}}_s,\, B^{\text{bw}}_s\}$ collects the scene-level BS configuration. 

A labeled fingerprint database is then constructed by collecting $N^{\text{data}}_s$ samples from $p_s$:
\begin{align}
\mathcal{D}_s = \big\{(\bm{H}_{s,i},\, \bm{p}_{s,i}^{\text{ue}},\, \bm{c}_s)\big\}_{i=1}^{N^{\text{data}}_s},
\label{eq:dataset}
\end{align}
where $(\bm{H}_{s,i},\, \bm{p}_{s,i}^{\text{ue}}) \sim p_s$, $\bm{H}_{s,i}$ is the $i$-th labeled CSI sample, $\bm{p}_{s,i}^{\text{ue}}$ is its associated UE position, and configuration $\bm{c}_s$ is shared by all samples within the scene.

\subsection{Problem Formulation}
\label{subsec:ProblemFormulation}

While the BS configuration $\bm{c}_s$ is typically available from site deployment records or operator-side metadata, the actual propagation environment $\mathcal{E}_s$ is difficult to acquire and analytically characterize in practice. To circumvent this difficulty, we adopt a retrieval-augmented formulation~\cite{huang2026retrieval}, in which a labeled fingerprint database collected in the same scene serves as an implicit proxy of its propagation environment. The defining property of this problem is that environment-specific information is encapsulated in the per-scene database $\mathcal{D}_s$ rather than in the model parameters $\{\bm{\theta}, \bm{\phi}\}$, so that, once trained on a sufficiently diverse set of scenes, the model can be deployed in a new scene by simply swapping $\mathcal{D}_s$ while $\{\bm{\theta}, \bm{\phi}\}$ remain fixed. This training-free cross-scene adaptation distinguishes our framework from conventional fingerprinting methods that require per-scene retraining.

Formally, given the $j$-th query CSI $\bm{H}_{s,j}^{\text{que}}$ collected at an unknown UE position $\bm{p}_{s,j}^{\text{ue,que}}$ in scene $s$, we factorize the localization task into two modules:
\begin{itemize}
    \item \textbf{Reference retrieval:} A retrieval module $\mathcal{G}_{\bm{\theta}}(\cdot)$, parameterized by $\bm{\theta}$, selects a subset of reference samples from $\mathcal{D}_s$ that are expected to be most informative for locating the query:
    \begin{align}
        \mathcal{D}_s^{\text{ret}}(\bm{H}_{s,j}^{\text{que}}) = \mathcal{G}_{\bm{\theta}}\big(\bm{H}_{s,j}^{\text{que}},\, \mathcal{D}_s\big),
        \label{eq:retrieval_func}
    \end{align}
    where each retrieved sample is a triplet $(\bm{H}_{s,k},\, \bm{p}_{s,k}^{\text{ue}},\, \bm{c}_s)$ inherited from $\mathcal{D}_s$, containing the original CSI, the corresponding UE position, and the scene configuration.

    \item \textbf{Context-augmented localization:} A localization network $\mathcal{F}_{\bm{\phi}}(\cdot)$, parameterized by $\bm{\phi}$, takes the query CSI together with the retrieved references as an augmented context, from which it predicts the query position:
    \begin{align}
        \hat{\bm{p}}_{s,j}^{\text{ue,que}} = \mathcal{F}_{\bm{\phi}}\big(\bm{H}_{s,j}^{\text{que}},\, \mathcal{D}_s^{\text{ret}}(\bm{H}_{s,j}^{\text{que}})\big).
        \label{eq:loc_func}
    \end{align}
\end{itemize}

To make $\{\bm{\theta}, \bm{\phi}\}$ scene-agnostic, the problem is formulated as minimizing the expected localization error over both the training scene distribution and the per-scene query distribution:
\begin{align}
\min_{\bm{\theta},\, \bm{\phi}} \quad & \mathbb{E}_{s \sim \mathcal{S}}\,\mathbb{E}_{(\bm{H}_{s,j}^{\text{que}},\, \bm{p}_{s,j}^{\text{ue,que}}) \sim p_s} \!\left[\,\big\|\bm{p}_{s,j}^{\text{ue,que}} - \hat{\bm{p}}_{s,j}^{\text{ue,que}}\big\|_2\,\right], \label{eq:obj} \\
\text{s.t.} \quad & \mathcal{D}_s^{\text{ret}}(\bm{H}_{s,j}^{\text{que}}) = \mathcal{G}_{\bm{\theta}}\big(\bm{H}_{s,j}^{\text{que}},\, \mathcal{D}_s\big), \label{eq:const_retrieval} \\
& \hat{\bm{p}}_{s,j}^{\text{ue,que}} = \mathcal{F}_{\bm{\phi}}\big(\bm{H}_{s,j}^{\text{que}},\, \mathcal{D}_s^{\text{ret}}(\bm{H}_{s,j}^{\text{que}})\big). \label{eq:const_loc}
\end{align}
The outer expectation over $\mathcal{S}$ means the parameters $\{\bm{\theta}, \bm{\phi}\}$ should capture properties common to all scenes rather than overfit to any particular one. However, solving the resulting optimization problem faces two key challenges. First, retrieving informative references from $\mathcal{D}_s$ needs a meaningful similarity criterion, the design of which poses the same challenge encountered in channel charting-based localization~\cite{stephan2024angle, palhares2025csi2vec, zhang2025unilocpro, huang2026retrieval}. Second, even with good references, predicting the query position from a small set of reference CSI samples and their positions is still hard, because the underlying channel-to-position mapping depends on the unobserved environment $\mathcal{E}_s$ and is not available in closed form. These two challenges motivate the FM-based channel representation and the ICL reasoning module developed in the following.

%====================================================================
\section{Proposed Method: RA-LWLM}
\label{sec:Method}
%====================================================================
In this section, we present the proposed RA-LWLM. We first give an overview of the overall framework, and then detail its three core components, i.e., the FM-based representation extraction, the retrieval module, and the MoE-based ICL localization module.

\subsection{Proposed RA-LWLM Framework}
\label{subsec:framework}

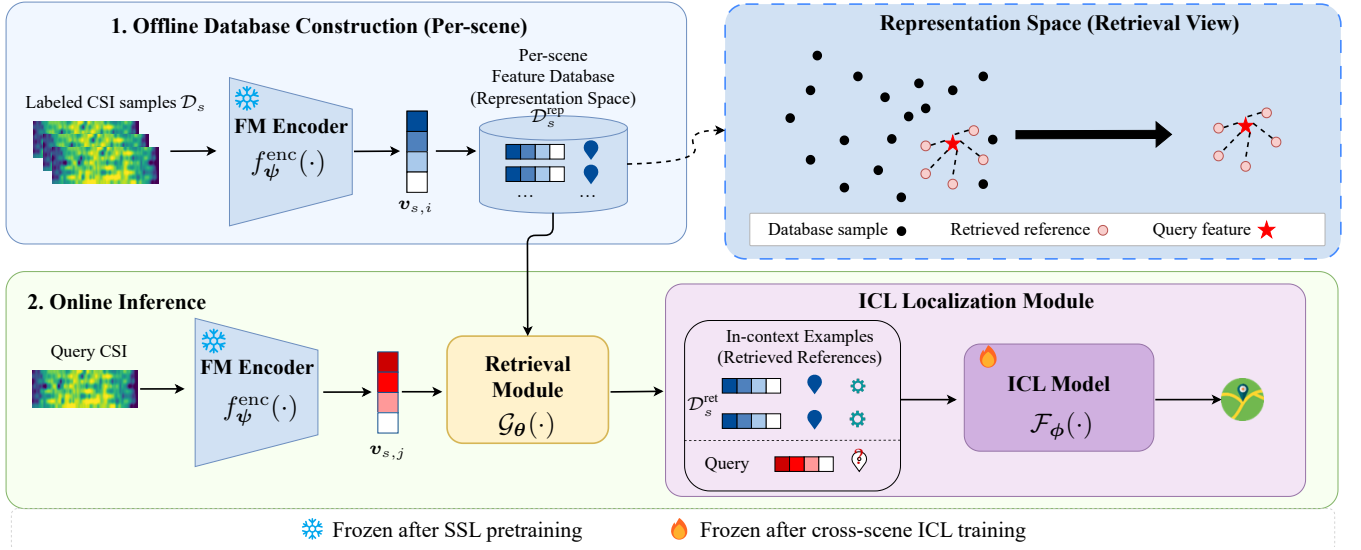
\begin{figure*}[t]
    \centering
    \begin{tikzpicture}
    \node (image) [anchor=south west]{\includegraphics[width=1.0\linewidth]{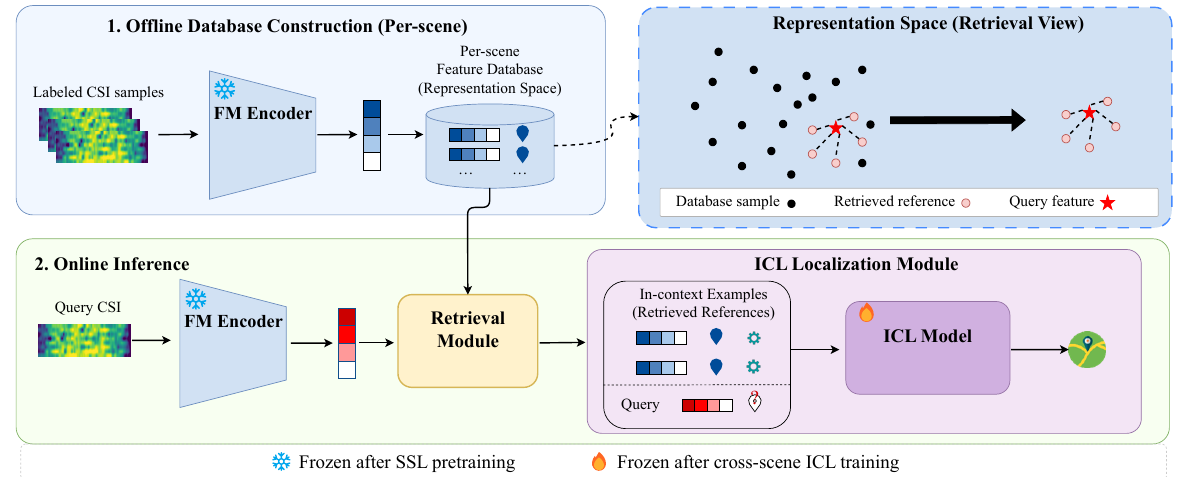}};
    \gettikzxy{(image.north east)}{\ix}{\iy};
    
    \node at (0.155*\ix,0.825*\iy)[rotate=0,anchor=north]{{\scriptsize $ \mathcal{D}_s$}};

    \node at (0.41*\ix,0.803*\iy)[rotate=0,anchor=north]{{\scriptsize $ \mathcal{D}_s^{\text{rep}}$}};

    \node at (0.315*\ix,0.640*\iy)[rotate=0,anchor=north]{{\scriptsize $ \bm{v}_{s,i}$}};
        
    \node at (0.22*\ix,0.73*\iy)[rotate=0,anchor=north]{{ $f^{\mathrm{enc}}_{\bm{\psi}}(\cdot
    )$}};

    \node at (0.20*\ix,0.300*\iy)[rotate=0,anchor=north]{{ $f^{\mathrm{enc}}_{\bm{\psi}}(\cdot
    )$}};

    \node at (0.295*\ix,0.21*\iy)[rotate=0,anchor=north]{{\scriptsize $ \bm{v}_{s,j}$}};

    \node at (0.395*\ix,0.27*\iy)[rotate=0,anchor=north]{{$\mathcal{G}_{\bm{\theta}}(\cdot)$}};

    \node at (0.522*\ix,0.30*\iy)[rotate=0,anchor=north]{{\scriptsize $ \mathcal{D}_s^{\text{ret}}$}};

    \node at (0.78*\ix,0.27*\iy)[rotate=0,anchor=north]{{$\mathcal{F}_{\bm{\phi}}(\cdot)$}};
    
    \end{tikzpicture}
    \caption{Overview of the RA-LWLM framework. Offline: $\mathcal{D}_s$ is encoded into a representation-space database $\mathcal{D}^{\text{rep}}_s$. Online: the encoded query retrieves the references from $\mathcal{D}^{\text{rep}}_s$, which are fused by the ICL module to predict the UE position. The FM Encoder is trained once via self-supervised learning and then kept frozen for feature database construction and query representation extraction. The ICL Localization Module is trained across multiple seen scenes and then kept frozen at inference, including on unseen scenes.}
    \label{fig:architecture}
    % \vspace{-5mm}
\end{figure*}

To address the two challenges identified above, the proposed RA-LWLM method consists of three main components: an FM-based channel representation model that extracts transferable CSI features, a retrieval module that searches the per-scene database to select informative references for each query, and an ICL localization module that fuses the query and the retrieved references into the final position estimate. The FM encoder builds a compact representation in which similarity search is both efficient and meaningful, addressing the first challenge, while the ICL module learns the channel-to-position mapping from a few retrieved examples without explicit environmental modeling, addressing the second. The overall architecture is illustrated in Fig.~\ref{fig:architecture}, and the three components are introduced as follows.
\begin{itemize}
    \item \textit{Foundation model encoder} $f^{\mathrm{enc}}_{\bm{\psi}}(\cdot)$: it maps each CSI matrix to a transferable representation vector. The encoder is pretrained via self-supervised learning on a large-scale unlabeled CSI dataset and is kept frozen throughout the downstream pipeline, serving as a universal channel representation that can be reused across scenes without modification \cite{pan2025large}.
    \item \textit{Retrieval module} $\mathcal{G}_{\bm{\theta}}$: given the encoded representation of the query, it searches the representation-space database of the current scene and returns a small set of references whose representations are closest to that of the query. Because the database is built in the encoder's representation space, the similarity search is scene-agnostic.
    \item \textit{ICL localization module} $\mathcal{F}_{\bm{\phi}}$: it takes the query CSI together with the retrieved references and performs in-context reasoning to produce the final UE position estimate. Inspired by the few-shot inference capability of LLMs, ICL conditions the prediction on a few in-context examples rather than updating any model parameters. In our setting, the retrieved references play the role of these examples and provide a scene-specific view of the CSI-to-position mapping. Therefore, the ICL module keeps $\bm{\phi}$ fixed across scenes and remains lightweight and training-free at deployment.
\end{itemize}

At runtime, the framework operates in two stages, as also indicated in Fig.~\ref{fig:architecture}. In the \emph{offline database construction} stage, all labeled CSI samples in $\mathcal{D}_s$ are passed through the frozen encoder once, and the resulting representation-space database $\mathcal{D}^{\text{rep}}_s$ (containing the encoded features, the corresponding UE positions, and the scene configuration) is stored on the BS side. This step has to be performed once per scene and is fully amortized over all subsequent online queries. In the \emph{online inference} stage, each incoming query CSI is encoded through the same encoder, the retrieval module fetches the similar references from $\mathcal{D}^{\text{rep}}_s$, and the ICL localization module predicts the UE position from the query-reference context.

\subsection{FM-based Representation Extraction}
\label{subsec:encoder}

The FM encoder provides the general channel representation on which both retrieval and ICL-based localization operate. We instantiate it with LWLM~\cite{pan2025large} and briefly describe below how a raw CSI observation is mapped into a representation vector.

\subsubsection{Input Preprocessing}
Given any CSI matrix $\bm{H}$ in scene $s$, whether a reference sample $\bm{H}_{s,i}$ from the database or a query $\bm{H}_{s,j}^{\text{que}}$, we decompose $\bm{H}$ into its magnitude and phase components before feeding it into the encoder. This decomposition is a bijective transformation of the original complex CSI and therefore preserves the full channel information:
\begin{align}
\bar{\bm{H}} = \big[\,|\bm{H}|;\; \angle \bm{H}\,\big] \in \mathbb{R}^{2 \times N^{\text{ant}}_s \times N^{\text{subc}}_s},
\label{eq:real_repr}
\end{align}
where the leading dimension of size $2$ stacks the amplitude and phase channels over the antenna-subcarrier domain.

\subsubsection{Encoder Architecture}
The encoder $f^{\mathrm{enc}}_{\bm{\psi}}(\cdot)$ follows the Transformer-based LWLM architecture~\cite{pan2025large}. Given the preprocessed input $\bar{\bm{H}}$, the LWLM encoder first divides it into $N_{\text{patch}}$ patches and maps each patch to an $N_{\text{embed}}$-dimensional embedding via a CNN patch-embedding layer, and prepends a learnable localization semantic token (LST) \cite{pan2025large} of the same dimension that aggregates the global channel semantics. This produces a tokenized sequence of length $N_{\text{patch}}+1$,
\begin{align}
\bm{X}^{\text{embed}} \,=\, \big[\bm{x}^{\text{embed}}_{0},\, \bm{x}^{\text{embed}}_{1},\, \dots,\, \bm{x}^{\text{embed}}_{N_{\text{patch}}}\big] + \bm{E}^{\text{seq}},
\label{eq:embed_seq}
\end{align}
where $\bm{x}^{\text{embed}}_{0} \in \mathbb{R}^{N_{\text{embed}}}$ denotes the LST token, $\{\bm{x}^{\text{embed}}_{n}\}_{n=1}^{N_{\text{patch}}}$ are the CNN patch embeddings, and $\bm{E}^{\text{seq}} \in \mathbb{R}^{(N_{\text{patch}}+1) \times N_{\text{embed}}}$ is the sinusoidal positional embedding matrix~\cite{vaswani2017attention} whose entries are
\begin{align}
\big[\bm{E}^{\text{seq}}\big]_{n,\, 2k} &\,=\, \sin\!\Big(n \,/\, 10000^{2k/N_{\text{embed}}}\Big), \label{eq:pe_sin}\\
\big[\bm{E}^{\text{seq}}\big]_{n,\, 2k+1} &\,=\, \cos\!\Big(n \,/\, 10000^{2k/N_{\text{embed}}}\Big), \label{eq:pe_cos}
\end{align}
with $n \in \{0,\dots,N_{\text{patch}}\}$ indexing the token position and $k$ indexing the embedding dimension. The embedded sequence $\bm{X}^{\text{embed}}$ then passes through $N_{\text{enc}}$ transformer encoder layers, and the encoder output is
\begin{align}
\bm{O} \,=\, f^{\mathrm{enc}}_{\bm{\psi}}(\bar{\bm{H}}) \,\in\, \mathbb{R}^{(N_{\text{patch}}+1) \times N_{\text{embed}}},
\label{eq:encoder_output}
\end{align}
where $\bm{O} = \big[\bm{o}_{0},\, \bm{o}_{1},\, \dots,\, \bm{o}_{N_{\text{patch}}}\big]$. Here $\bm{o}_{0}$ is the LST token that summarizes the global channel semantics, and $\bm{o}_{1},\,\dots,\,\bm{o}_{N_{\text{patch}}}$ are the patch tokens that preserve the local spatial-frequency feature. Because the patch embedding is convolutional and the sinusoidal positional encoding is constructed on the fly from the input shape, the same encoder admits inputs of varying $N^{\text{ant}}_s$ and $N^{\text{subc}}_s$ across scenes.

\subsubsection{Feature Extraction}
Since the encoder output $\bm{O}$ is still high-dimensional and not directly suitable as a retrieval key, we distill it into two complementary representations, i.e., $
\bm{z}^{\mathrm{lst}} = \bm{o}_{0}$ and $
\bm{z}^{\mathrm{mp}} = \tfrac{1}{N_{\text{patch}}}\sum_{n=1}^{N_{\text{patch}}} \bm{o}_{n} $,
where $\bm{z}^{\mathrm{lst}}$ is the LST token capturing the global channel semantics, while $\bm{z}^{\mathrm{mp}}$ mean-pools the patch tokens to summarize the local spatial-frequency details. These two are concatenated into the final representation vector
\begin{align}
\bm{v} = \big[\,\bm{z}^{\mathrm{lst}};\; \bm{z}^{\mathrm{mp}}\,\big] \in \mathbb{R}^{2 N_{\text{embed}}}.
\label{eq:feature}
\end{align}
Applying this extraction pipeline to a reference CSI $\bm{H}_{s,i}$ and to a query CSI $\bm{H}_{s,j}^{\text{que}}$ yields the reference representation $\bm{v}_{s,i}$ and the query representation $\bm{v}_{s,j}^{\text{que}}$, respectively.

\subsection{Retrieval Module Design}
\label{subsec:retrieval}

For each online query, the retrieval module $\mathcal{G}_{\bm{\theta}}$ selects a small set of references from the per-scene database that are then fed into the ICL localization module. It consists of two steps, namely an offline representation-space database construction performed once per scene and an online similarity search executed for each incoming query. In this work, $\mathcal{G}_{\bm{\theta}}$ contains no trainable parameters, and its behavior is fully determined by the representation space induced by the frozen FM encoder $f^{\mathrm{enc}}_{\bm{\psi}}$. We retain the parameterized notation $\mathcal{G}_{\bm{\theta}}$ for two reasons. First, the quality of the retrieved references is ultimately governed by the FM pretraining that shapes this representation space. Second, the notation leaves room for future extensions in which the retrieval mapping itself becomes trainable.

\subsubsection{Representation-Space Database Construction}
For each scene $s$, we pass every labeled CSI sample in $\mathcal{D}_s$ through the frozen FM encoder once to obtain its representation, and cache the resulting representation-space database
\begin{align}
\mathcal{D}^{\text{rep}}_s \,=\, \big\{(\bm{v}_{s,i},\; \bm{p}_{s,i}^{\text{ue}},\; \bm{c}_s)\big\}_{i=1}^{N^{\text{data}}_s},
\label{eq:feat_db}
\end{align}
where $\bm{v}_{s,i}$ is obtained from~\eqref{eq:feature}. For each scene, this step is performed once and does not involve any trainable parameters. Critically, adapting the system to a new scene $s'$ only requires running the frozen encoder over the labeled CSIs in that scene to build $\mathcal{D}^{\text{rep}}_{s'}$, so that no parameter update is needed. Moreover, when the propagation environment of an existing scene drifts over time, the system is maintained simply by refreshing the corresponding $\mathcal{D}^{\text{rep}}_{s'}$ with newly collected labeled CSIs, without retraining either the FM encoder or the ICL module.

\subsubsection{Similarity-Based Retrieval}
At inference, given the $j$-th query CSI $\bm{H}_{s,j}^{\text{que}}$ in scene $s$ with its encoder representation $\bm{v}_{s,j}^{\text{que}}$, the retrieval module returns the top-$K$ nearest references from $\mathcal{D}^{\text{rep}}_s$, measured by the Euclidean distance on the encoder representations. The retrieved reference set is
\begin{align}
\mathcal{D}_s^{\text{ret}}(\bm{H}_{s,j}^{\text{que}}) \,=\, \big\{(\bm{v}_{s,k},\, \bm{p}_{s,k}^{\text{ue}},\, \bm{c}_s)\big\}_{k=1}^{K},
\label{eq:retrieval_set}
\end{align}
where the $K$ entries are drawn from $\mathcal{D}^{\text{rep}}_s$ and re-indexed in ascending order of distance to the query, i.e.,
\begin{align}
\big\|\bm{v}_{s,j}^{\text{que}} - \bm{v}_{s,1}\big\|_2 &\le \dots \le \big\|\bm{v}_{s,j}^{\text{que}} - \bm{v}_{s,K}\big\|_2 \nonumber \\
&\le \!\!\min_{i \notin \{1,\dots,K\}}\! \big\|\bm{v}_{s,j}^{\text{que}} - \bm{v}_{s,i}\big\|_2.
\label{eq:retrieval_order}
\end{align}
In addition, the retrieval module computes a softmax-normalized weight for each retrieved reference,
\begin{align}
w_{s,j,k} \,=\, \frac{\exp\!\big(-\big\|\bm{v}_{s,j}^{\text{que}} - \bm{v}_{s,k}\big\|_2\big)}{\sum_{l=1}^{K} \exp\!\big(-\big\|\bm{v}_{s,j}^{\text{que}} - \bm{v}_{s,l}\big\|_2\big)},
\label{eq:retrieval_weight}
\end{align}
which will be used as a soft prior over the references in the downstream ICL module.

% During training, the query sample's own database entry is excluded from the candidate set via leave-one-out to prevent trivial retrieval. The retrieval module is therefore parameter-free (i.e., $\bm{\theta} = \varnothing$); the per-query selection of how many of the $K_{\max}$ retrieved references are actually used is deferred to a multi-expert mechanism inside the ICL localization module described in Section~\ref{subsec:icl_module}.

\subsection{MoE-Based ICL Localization}
\label{subsec:icl_module}

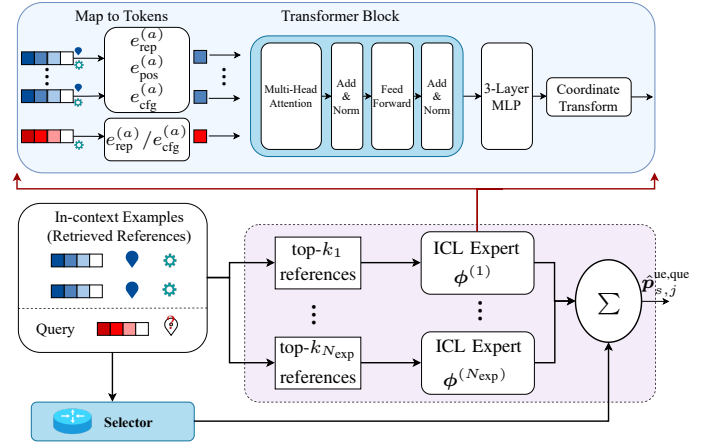
\begin{figure}[t]
\centering
   \begin{tikzpicture}
    \node (image) [anchor=south west]{\includegraphics[width=1.0\linewidth]{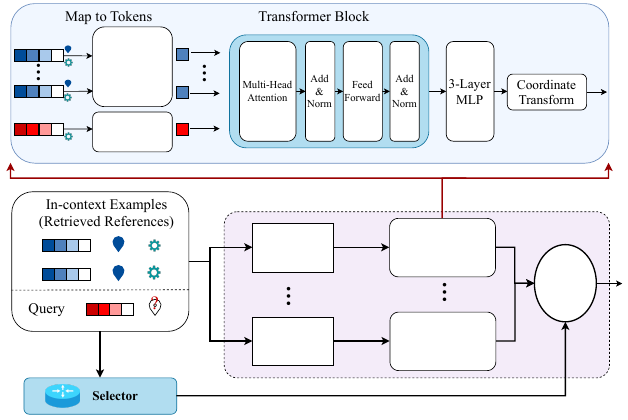}};
    \gettikzxy{(image.north east)}{\ix}{\iy};
    
    \node at (0.462*\ix,0.48*\iy)[rotate=0,anchor=north]{\scriptsize { top-$ k_1$}};
    \node at (0.463*\ix,0.425*\iy)[rotate=0,anchor=north]{\scriptsize { references}};

    \node at (0.463*\ix,0.27*\iy)[rotate=0,anchor=north]{\scriptsize { top-$ k_{N_\text{exp}}$}};
    \node at (0.463*\ix,0.205*\iy)[rotate=0,anchor=north]{\scriptsize { references}};

     \node at (0.693*\ix,0.48*\iy)[rotate=0,anchor=north]{\scriptsize { ICL Expert }};
        \node at (0.693*\ix,0.43*\iy)[rotate=0,anchor=north]{\scriptsize {$\bm{\phi}^{(1)}$}};

    \node at (0.696*\ix,0.27*\iy)[rotate=0,anchor=north]{\scriptsize { ICL Expert }};
    \node at (0.696*\ix,0.21*\iy)[rotate=0,anchor=north]{\scriptsize {$\bm{\phi}^{(N_\text{exp})}$}};

    \node at (0.22*\ix,0.938*\iy)[rotate=0,anchor=north]{{\scriptsize $ e_{\text{rep}}^{(a)} $}};

    \node at (0.22*\ix,0.879*\iy)[rotate=0,anchor=north]{{\scriptsize $ e_{\text{pos}}^{(a)} $}};

    \node at (0.22*\ix,0.820*\iy)[rotate=0,anchor=north]{{\scriptsize $ e_{\text{cfg}}^{(a)}$}};

     \node at (0.219*\ix,0.730*\iy)[rotate=0,anchor=north]{{\scriptsize $ e_{\text{rep}}^{(a)} / e_{\text{cfg}}^{(a)}$}};

     \node at (0.89*\ix,0.370*\iy)[rotate=0,anchor=north]{{\footnotesize $\sum$}};

     \node at (0.973*\ix,0.414*\iy)[rotate=0,anchor=north]{{\scriptsize $\hat{\bm{p}}_{s,j}^{\text{ue,que}}$}};
    
    \end{tikzpicture}
\caption{Overview of the ICL localization module. Both the selector and each of the $N_{\text{exp}}$ experts are trainable modules.}
\label{fig:ICL}
\end{figure}

The optimal context size for ICL inference is inherently query-dependent. When the query falls in a region densely sampled by the database, a small number of top references is already informative, whereas in sparse or NLOS regions, more references are needed to average out retrieval noise. Moreover, using too few references may render the ICL inference unstable, as the prediction relies heavily on the geometry of a single nearest neighbor and is sensitive to retrieval noise. Conversely, using too many references inevitably brings in distant or weakly correlated samples, which dilute the informative signal and bias the prediction away from the true UE position. To address this, the ICL localization module $\mathcal{F}_{\bm{\phi}}$ adopts an MoE design that consists of $N_{\text{exp}}$ ICL experts and a learnable selector, with learnable parameters $\bm{\phi}$ comprising the per-expert weights $\{\bm{\phi}^{(a)}\}_{a=1}^{N_{\text{exp}}}$ and the selector weights $\bm{\phi}^{\text{sel}}$. A single expert operating at a fixed context size therefore cannot perform uniformly well across all queries. To exploit this, we define a discrete set of $N_{\text{exp}}$ candidate context sizes $\mathcal{K} = \{k_1, \dots, k_{N_{\text{exp}}}\}$ with $k_1 < k_2 < \cdots < k_{N_{\text{exp}}} \le K$, covering a range of context sizes from a few nearest references to the full retrieval budget. The $a$-th expert is specialized at the context size $k_a \in \mathcal{K}$, and uses only the top-$k_a$ references from $\mathcal{D}_s^{\text{ret}}(\bm{H}_{s,j}^{\text{que}})$ to produce a per-expert position estimate. The selector inspects the retrieval output and emits a routing distribution over the $N_{\text{exp}}$ experts, and the final UE position estimate is a soft mixture of the per-expert predictions. We describe the two parts below. The schematic diagram of the ICL localization module is shown in Fig.~\ref{fig:ICL}.

\subsubsection{Per-Expert ICL Inference}
All $N_{\text{exp}}$ experts share the same model architecture but have independent parameters, so that each expert can be trained to specialize in its assigned context size $k_a$ without interfering with the others. For the $a$-th expert, the parameter set $\bm{\phi}^{(a)}$ comprises three input-embedding layers $\{e_{\text{rep}}^{(a)}, e_{\text{pos}}^{(a)}, e_{\text{cfg}}^{(a)}\}$ that map the retrieved representation, the relative reference position, and the scene configuration\footnote{Although $\bm{c}_s$ takes the same value across all reference and query tokens within a scene, we still inject it into every token because we expect each ICL token to learn a unified, configuration-aware representation, which paves the way for future studies on robust localization under varying BS configurations within a single scene.} into the ICL token space, an $N_{\text{enc}}^{\text{ICL}}$-layer transformer encoder that performs in-context reasoning, and an output MLP $e_{\text{pos}}^{\text{out},(a)}$ that decodes the post-transformer query state into a position residual. Specifically, we first renormalize the retrieval weights from~\eqref{eq:retrieval_weight} over its top-$k_a$ references,
\begin{align}
\tilde{w}_{s,j,k}^{(a)} \,=\, w_{s,j,k} \,/\, \textstyle\sum_{l=1}^{k_a} w_{s,j,l}, \quad k = 1, \dots, k_a,
\label{eq:expert_weight}
\end{align}
and form an expert-specific weighted centroid of the retrieved positions,
\begin{align}
\bar{\bm{p}}_{s,j}^{(a)} \,=\, \sum_{k=1}^{k_a} \tilde{w}_{s,j,k}^{(a)} \,\bm{p}_{s,k}^{\text{ue}}.
\label{eq:expert_centroid}
\end{align}
Each retrieved position is then re-expressed relative to this centroid and rescaled by a spatial normalization scale $\sigma_{\text{pos}}$,
\begin{align}
\tilde{\bm{p}}_{s,j,k}^{(a)} \,=\, (\bm{p}_{s,k}^{\text{ue}} - \bar{\bm{p}}_{s,j}^{(a)}) \,/\, \sigma_{\text{pos}},
\label{eq:expert_relpos}
\end{align}
which removes the dependence on absolute coordinates so that the downstream ICL transformer learns spatial patterns among the retrieved references rather than memorizing absolute positions specific to any training scene.

The $k$-th reference is then tokenized into a common ICL embedding space $\mathbb{R}^{N_{\text{ICL}}}$ as
\begin{align}
\bm{t}_{s,j,k}^{\text{ref},(a)} \,=\, e_{\text{rep}}^{(a)}(\bm{v}_{s,k}) \,+\, e_{\text{pos}}^{(a)}\!\big(\tilde{\bm{p}}_{s,j,k}^{(a)}\big) \,+\, e_{\text{cfg}}^{(a)}(\bm{c}_s),
\label{eq:expert_ref_token}
\end{align}
where $e_{\text{rep}}^{(a)}, e_{\text{pos}}^{(a)}, e_{\text{cfg}}^{(a)}$ are MLPs with expert-specific parameters. The query token is constructed analogously but without a position term,
\begin{align}
\bm{t}_{s,j}^{\text{que},(a)} \,=\, e_{\text{rep}}^{(a)}(\bm{v}_{s,j}^{\text{que}}) \,+\, e_{\text{cfg}}^{(a)}(\bm{c}_s).
\label{eq:expert_query_token}
\end{align}
The reference tokens $\{\bm{t}_{s,j,k}^{\text{ref},(a)}\}_{k=1}^{k_a}$ and the query token $\bm{t}_{s,j}^{\text{que},(a)}$ are stacked into an input sequence of length $k_a+1$ and processed by $N_{\text{enc}}^{\text{ICL}}$ transformer encoder layers, producing the post-transformer query state $\bm{t}_{s,j}^{\text{que},\star,(a)}$. A 3-layer MLP $e_{\text{pos}}^{\text{out},(a)}$ then maps this query state to a centered-normalized residual, and the per-expert UE position estimate is obtained by combining the residual with the expert centroid and the normalization scale,
\begin{align}
\hat{\bm{p}}_{s,j}^{(a)} \,=\, \bar{\bm{p}}_{s,j}^{(a)} \,+\, \sigma_{\text{pos}} \cdot e_{\text{pos}}^{\text{out},(a)}\!\big(\bm{t}_{s,j}^{\text{que},\star,(a)}\big).
\label{eq:expert_pred}
\end{align}
Based on the position centering and normalization in~\eqref{eq:expert_relpos}, each expert only needs to learn a small correction on top of its weighted-KNN starting point, which stabilizes training.

\subsubsection{Selector and Multi-Expert Routing}
The selector $g_{\bm{\phi}^{\text{sel}}}$ is a 3-layer MLP that maps the query, the retrieved references, and their retrieval weights to $N_{\text{exp}}$ expert logits. Concretely, it takes as input the query representation $\bm{v}_{s,j}^{\text{que}}$, the scene configuration $\bm{c}_s$, the retrieved reference representations $\{\bm{v}_{s,k}\}_{k=1}^{K}$, and the retrieval weights $\{w_{s,j,k}\}_{k=1}^{K}$. Denoting this input collection by $\bm{f}_{s,j}^{\text{sel}}$, the selector produces a vector of routing scores $\boldsymbol{\ell}_{s,j} = g_{\bm{\phi}^{\text{sel}}}(\bm{f}_{s,j}^{\text{sel}}) \in \mathbb{R}^{N_{\text{exp}}}$, where $\boldsymbol{\ell}_{s,j} = [\ell_{s,j}^{(1)}, \dots, \ell_{s,j}^{(N_{\text{exp}})}]^{\top}$ and the $a$-th entry $\ell_{s,j}^{(a)}$ corresponds to the raw routing score of the $a$-th expert before normalization. The selector scores are converted into a routing distribution via a softmax:
\begin{align}
\pi_{s,j}^{(a)} \,=\, \frac{\exp(\ell_{s,j}^{(a)})}{\sum_{a'=1}^{N_{\text{exp}}} \exp(\ell_{s,j}^{(a')})}, \quad a = 1, \dots, N_{\text{exp}}.
\label{eq:gate}
\end{align}
The final UE position estimate is then a soft mixture over the per-expert predictions,
\begin{align}
\hat{\bm{p}}_{s,j}^{\text{ue,que}} \,=\, \sum_{a=1}^{N_{\text{exp}}} \pi_{s,j}^{(a)} \,\hat{\bm{p}}_{s,j}^{(a)}.
\label{eq:moe_pred}
\end{align}
In practice, the soft-mixture mode in~\eqref{eq:moe_pred} requires $N_{\text{exp}}$ expert forward passes per query to fully exploit the specialization of all experts. When inference latency is a concern, a hard top-$1$ routing $\hat{\bm{p}}_{s,j}^{\text{ue,que}} = \hat{\bm{p}}_{s,j}^{(a^\star)}$ with $a^\star = \arg\max_a \pi_{s,j}^{(a)}$ can be used instead, reducing the cost to a single expert forward pass at the price of a small accuracy loss.

\section{Training Details}
\label{subsec:training}

RA-LWLM is trained in two stages: (i) self-supervised pretraining of the FM encoder, (ii) supervised end-to-end training of the retrieval-augmented localization.

\subsection{FM Pretraining}
The FM encoder is pretrained on large-scale unlabeled CSI data using a single domain-transformation invariance (DTI)-based self-supervised learning method~\cite{pan2025large}.\footnote{Other self-supervised learning methods can also be used to pretrain the FM encoder. Since the choice of pretraining objective is not the focus of this work, we adopt the DTI-based method for its simplicity and effectiveness. A systematic study of the optimal self-supervised learning objective for retrieval-based localization is left for future work.}

 DTI encourages the encoder to extract cross-domain consistent features by learning the transformation between two physically meaningful channel domains: the spatial-frequency domain and the angle-delay domain. Based on DTI, the pretrained features retain the essential channel information that is invariant to such domain transformation. Specifically, given the spatial-frequency CSI $\bm{H}_{s,i}$, its angle-delay expression is obtained via a 2D-DFT,
$\bm{H}_{s,i}^{\text{DTI}} \,=\, \bm{W}_\theta^H \,\bm{H}_{s,i}\, \bm{W}_\tau^{*}$
where $\bm{W}_\theta \in \mathbb{C}^{N_s^{\text{ant}} \times N_s^{\text{ant}}}$ and $\bm{W}_\tau \in \mathbb{C}^{N_s^{\text{subc}} \times N_s^{\text{subc}}}$ are unitary DFT matrices that map the spatial domain to the angle domain and the frequency domain to the delay domain, respectively, with entries $[\bm{W}_\theta]_{i_1, i_2} = \tfrac{1}{\sqrt{N_s^{\text{ant}}}}\, e^{-j 2\pi i_1 i_2 / N_s^{\text{ant}}}$ and $[\bm{W}_\tau]_{i_1, i_2} = \tfrac{1}{\sqrt{N_s^{\text{subc}}}}\, e^{-j 2\pi i_1 i_2 / N_s^{\text{subc}}}$.
During pretraining, the encoder $f^{\mathrm{enc}}_{\bm{\psi}}(\cdot)$ first maps $\bm{H}_{s,i}$ into a latent representation, which is then decoded to the angle-delay view by a lightweight DTI decoder $f^{\mathrm{DTI}}_{\bm{\psi}^{\text{DTI}}}(\cdot)$,
\begin{align}
\hat{\bm{H}}_{s,i}^{\text{DTI}} \,=\, f^{\mathrm{DTI}}_{\bm{\psi}^{\text{DTI}}}\!\Big(f^{\mathrm{enc}}_{\bm{\psi}}(\bm{H}_{s,i})\Big).
\label{eq:dti_decode}
\end{align}
The FM encoder and the DTI decoder are jointly optimized by minimizing the loss between $\hat{\bm{H}}_{s,i}^{\text{DTI}}$ and $\bm{H}_{s,i}^{\text{DTI}}$ \cite{pan2025large},
\begin{align}
\mathcal{L}_{\text{DTI}} \,=\, \frac{1}{N_{\text{bat}}}\!\!\sum_{\bm{H}_{s,i}\in\mathcal{N}_{\text{bat}}}\!\!\left[\,1 \,-\, \frac{\big|\big\langle \bm{H}_{s,i}^{\text{DTI}},\, \hat{\bm{H}}_{s,i}^{\text{DTI}} \big\rangle\big|}{\big\|\bm{H}_{s,i}^{\text{DTI}}\big\| \cdot \big\|\hat{\bm{H}}_{s,i}^{\text{DTI}}\big\|}\,\right],
\label{eq:loss_dti}
\end{align}
where $\mathcal{N}_{\text{bat}}$ is a mini-batch of $N_{\text{bat}}$ training samples drawn from different scenes and different positions within each scene. After pretraining, the DTI decoder is discarded and only the encoder $f^{\mathrm{enc}}_{\bm{\psi}}(\cdot)$ is retained for downstream use.

\subsection{Retrieval-Augmented ICL Training}

With the FM encoder frozen, the ICL localization parameters $\bm{\phi}$, comprising the per-expert ICL weights $\bm{\phi}^{(a)}$ and the selector weights $\bm{\phi}^{\text{sel}}$, are trained on labeled CSI samples drawn from a set of training scenes $\mathcal{S}_{\text{train}}$, each containing $N^{\text{data}}_s$ labeled CSI-position pairs. We adopt a two-stage procedure that first equips each expert with a strong specialized initialization at its fixed context size, and then learns a query-dependent routing on top of these frozen experts. 

\textbf{Stage 1: Per-Expert Pretraining.}
For each $a \in \{1,\dots,N_{\text{exp}}\}$, we train the $a$-th expert in isolation at its fixed context size $k_a$. At each iteration, we sample a mini-batch from the union of per-scene databases $\bigcup_{s \in \mathcal{S}_{\text{train}}} \mathcal{D}_s$ and treat each sampled pair $(\bm{H}_{s,j},\,\bm{p}_{s,j}^{\text{ue}})$ as a query. To prevent trivial retrieval, the query's own entry is excluded from the candidate set via leave-one-out, so that the retrieval module searches only over $\mathcal{D}_s \setminus \{(\bm{H}_{s,j},\,\bm{p}_{s,j}^{\text{ue}})\}$. The module returns the top-$K$ references, and the $a$-th expert keeps only its top-$k_a$ subset to produce $\hat{\bm{p}}_{s,j}^{(a)}$ as in~\eqref{eq:expert_pred}. Each expert is supervised by the scale-normalized localization loss
\begin{align}
\mathcal{L}_{\text{loc}}\!\big(\bm{\phi}^{(a)}\big) \,=\, \frac{1}{|\mathcal{S}_{\text{train}}|} \sum_{s \in \mathcal{S}_{\text{train}}} \frac{1}{N^{\text{data}}_s}\sum_{j=1}^{N^{\text{data}}_s} \big\|\bm{p}_{s,j}^{\text{ue}} - \hat{\bm{p}}_{s,j}^{(a)}\big\|_2,
\label{eq:loss_expert}
\end{align}
where $\hat{\bm{p}}_{s,j}^{(a)}$ is the per-expert position estimate produced by the $a$-th expert from its top-$k_a$ retrieved references, as defined in~\eqref{eq:expert_pred}. Optimizing~\eqref{eq:loss_expert} yields $N_{\text{exp}}$ specialized expert weights $\{\bm{\phi}^{(a)}\}_{a=1}^{N_{\text{exp}}}$, each tuned to predict from exactly $k_a$ retrieved neighbors.

\textbf{Stage 2: Joint Routing Training.}
We then freeze the $N_{\text{exp}}$ pretrained experts and train only the selector $g_{\bm{\phi}^{\text{sel}}}$, against the same scale-normalized loss applied this time to the soft-mixed prediction in~\eqref{eq:moe_pred},
\begin{align}
\mathcal{L}\!\big(\bm{\phi}^{\text{sel}}\big) \,=\, \frac{1}{|\mathcal{S}_{\text{train}}|} \sum_{s \in \mathcal{S}_{\text{train}}} \frac{1}{N^{\text{data}}_s} \sum_{j=1}^{N^{\text{data}}_s} \big\|\bm{p}_{s,j}^{\text{ue}} - \hat{\bm{p}}_{s,j}^{\text{ue}}\big\|_2,
\label{eq:loss_loc}
\end{align}
Freezing the experts during Stage~2 prevents the selector from collapsing onto a single expert and degrading the others, and yields a balanced routing distribution that preserves the specialization of all $N_{\text{exp}}$ experts.

After the two stages, the trained model, comprising the frozen FM encoder $\bm{\psi}$, the per-expert parameters $\{\bm{\phi}^{(a)}\}_{a=1}^{N_{\text{exp}}}$, and the selector $\bm{\phi}^{\text{sel}}$, is reused across all deployment scenes. Adaptation to a new scene is achieved purely by swapping the per-scene database, so RA-LWLM keeps a lightweight training process and achieves training-free cross-scene adaptation by design.

\subsection{Model Implementation Details}

We summarize the model architecture of RA-LWLM as follows. The FM encoder follows the LWLM design~\cite{pan2025large}. The CNN patch-embedding layer is a single 2D convolution with kernel size $K_{\text{cnn}} = 4$ and stride $S_{\text{cnn}} = 4$, which slides over the antenna-subcarrier domain of the preprocessed CSI input $\bar{\bm{H}}$ and produces $
N_{\text{patch}} \,=\, \left\lfloor \frac{N^{\text{ant}}_s - K_{\text{cnn}}}{S_{\text{cnn}}} + 1 \right\rfloor \times \left\lfloor \frac{N^{\text{subc}}_s - K_{\text{cnn}}}{S_{\text{cnn}}} + 1 \right\rfloor $
non-overlapping patches, each mapped to an $N_{\text{embed}} = 256$ dimensional token. A learnable LST token of the same dimension is prepended to aggregate global channel semantics, and the resulting $(N_{\text{patch}}+1)$ tokens are fed into $N_{\text{enc}} = 4$ Transformer encoder layers with $4$ attention heads, producing the encoder output $\bm{O}$. The final channel representation $\bm{v}$ is then extracted from $\bm{O}$ by concatenating the LST token with the mean-pooled patch tokens, and is shared by both the retrieval module and the ICL module.

For the ICL module, all $N_{\text{exp}}$ experts share the same architecture but maintain independent parameters $\bm{\phi}^{(a)}$. At each inference, the retrieval module returns the top $K = 15$ references, and the ICL module routes them through $N_{\text{exp}} = 5$ experts with per-expert context sizes $\mathcal{K} = \{3,\,6,\,9,\,12,\,15\}$. For each expert, the three input-embedding MLPs $e_{\text{rep}}^{(a)}, e_{\text{pos}}^{(a)}, e_{\text{cfg}}^{(a)}$ are 2-layer feedforward networks with hidden dimension $128$, projecting their inputs into the common ICL token space of dimension $N_{\text{ICL}} = 256$. The resulting reference and query tokens are processed by a Transformer encoder with $N_{\text{enc}}^{\text{ICL}} = 4$ layers and $4$ attention heads. The output MLP $e_{\text{pos}}^{\text{out},(a)}$ is a 3-layer feedforward network with hidden dimensions $256$ and $128$, which decodes the post-transformer query state into the position residual that yields the per-expert position estimate in~\eqref{eq:expert_pred}. The selector $g_{\bm{\phi}^{\text{sel}}}$ is a 3-layer feedforward MLP with hidden dimensions $512$ and $256$, and outputs a vector of $N_{\text{exp}}$ routing logits that are converted into the soft routing distribution by the softmax.

\color{black}

%====================================================================
\section{Simulation Results}
\label{sec:Experiments}
%====================================================================

\subsection{Experiment Settings}
\label{subsec:settings}

\begin{figure}[t]
\centering
\includegraphics[width=\linewidth]{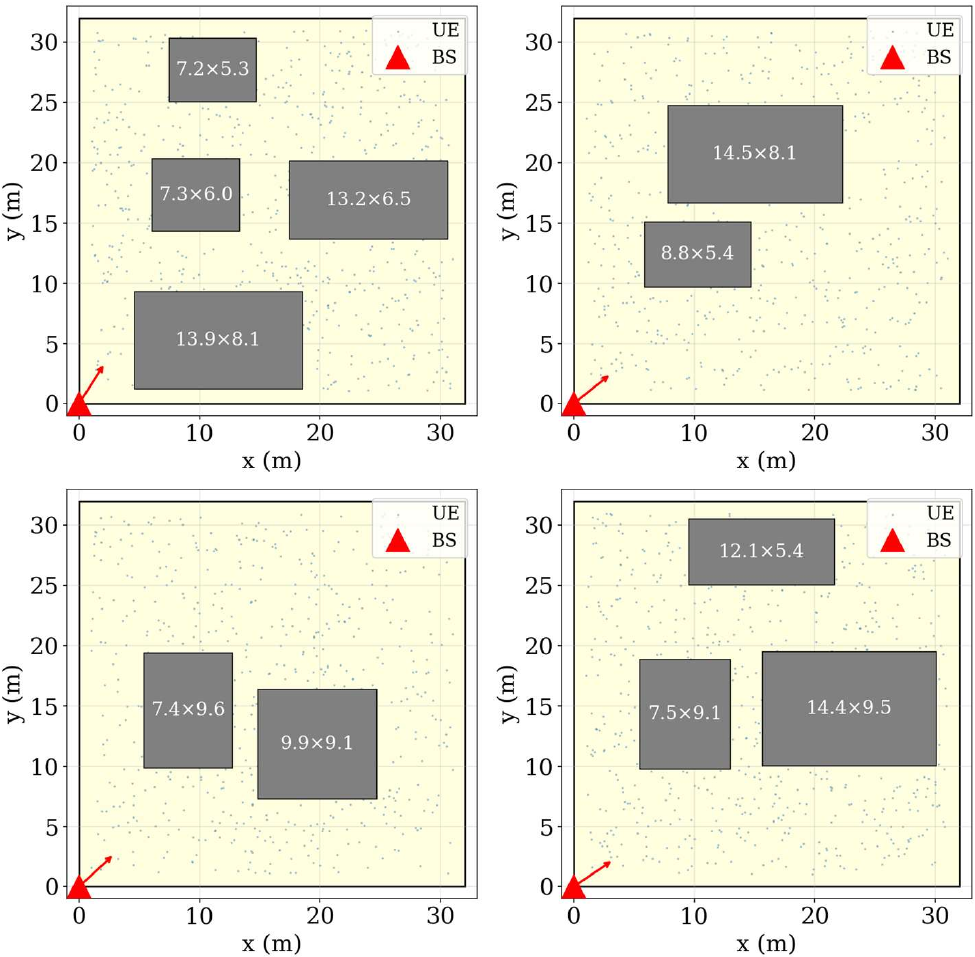}
\caption{Top-down visualization of 4 representative scenes drawn from the random scene-generation procedure. The BS is placed at the origin (red triangle), with its ULA boresight indicated by the red arrow. Grey rectangles denote buildings. The 4 scenes differ in BS height, orientation, number and size of buildings, and BS configuration, demonstrating the heterogeneity of the deployment conditions considered in our evaluation.}
\label{fig:scene_examples}
\end{figure}

We use the Sionna~\cite{Sionna} ray-tracing simulator to generate channel data. For each scene $s$, the BS-side configuration $\bm{c}_s$ and the surrounding propagation environment $\mathcal{E}_s$ are randomly generated. Specifically, the BS antenna height is sampled as $z_s^{\text{bs}} \sim \mathcal{U}(15.0,\,20.0)$~m, the bandwidth as $B^{\text{bw}}_s \sim \mathcal{U}\{5,\,10,\,20\}$~MHz, the number of BS antennas as $N^{\text{ant}}_s \sim \mathcal{U}\{8,\,16,\,32\}$, and the BS orientation as $\varphi^{\text{az}}_s \sim \mathcal{U}(25.0^\circ,\,65.0^\circ)$. The carrier frequency is fixed at $3.5$~GHz, and the number of subcarriers is fixed at $N^{\text{subc}}_s = 128$ across all scenes. Each scene additionally contains a number of randomly placed rectangular concrete buildings drawn uniformly from $\{2,3,4\}$, with a fixed height of $10$~m and lengths and widths sampled from $\mathcal{U}(5,\,16)$~m and $\mathcal{U}(5,\,10)$~m, respectively. Unless otherwise specified, for each scene, $N^{\text{data}}_s = 4000$ randomly sampled CSI-position pairs serve as the fingerprint database $\mathcal{D}_s$, and another $1000$ samples form the test set. We train and evaluate RA-LWLM and all baselines under two settings, i.e., (i) \emph{seen scenes} (SS), where 20 scenes are used for training and the test samples are drawn at new UE positions within these 20 training scenes, measuring in-scene generalization to new positions within a trained propagation environment, and (ii) \emph{unseen scenes} (US), where the test samples come from an additional 10 scenes, measuring cross-scene generalization to entirely new environments and BS configurations. Fig.~\ref{fig:scene_examples} visualizes 4 representative scenes generated by the above procedure, illustrating the diversity of BS configurations and propagation environments.

% \begin{figure}[t]
% \centering
% \includestandalone[scale=0.80]{Figures_Tikz/ret_results}
%     \caption{Mean physical distance between the query and its top-$K$ retrieved references under different retrieval strategies.}
% \label{fig:retrieval_quality}
% \end{figure}

\begin{figure}[t]
    \centering
    {\includegraphics[scale=0.9]{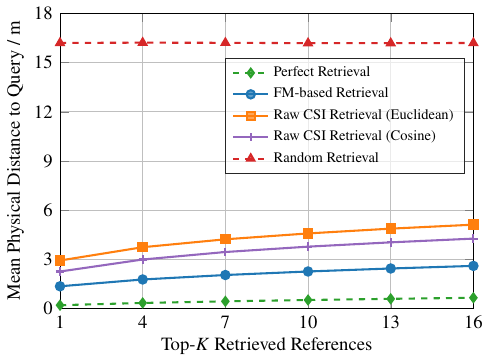}}
    \caption{Mean physical distance between the query and its top-$K$ retrieved references under different retrieval strategies.}
\label{fig:retrieval_quality}
\end{figure}

\subsection{Experiment Results}

\subsubsection{Analysis of Retrieval Quality}

This experiment isolates the quality of the retrieval module, which is the prerequisite for any retrieval-augmented inference to work. Rather than evaluating the final localization accuracy, we measure the mean physical distance between the query and its top-$K$ retrieved references. This metric largely reflects retrieval quality, since references physically close to the query are more likely to share similar propagation conditions and thus to be informative for the subsequent ICL inference. Fig.~\ref{fig:retrieval_quality} reports this metric averaged over the unseen test scenes. We compare four retrieval strategies. (i) \textbf{Perfect Retrieval} assumes direct access to the ground-truth UE positions of all database samples and directly retrieves the $K$ samples physically closest to the query, serving as the theoretical lower bound. (ii) \textbf{FM-based Retrieval} is our proposed strategy that uses the FM-encoded representation. (iii) \textbf{Raw CSI Retrieval} directly searches the raw CSI, which is conceptually similar to channel charting~\cite{studer2018channel} that uses the same channel-domain similarity to train its low-dimensional embedding. We report both the Euclidean distance and the cosine similarity as the search metric \cite{stephan2024angle}. (iv) \textbf{Random Retrieval} uniformly samples $K$ references from the database and serves as the reference upper bound. With $K = 1$, Perfect Retrieval achieves 0.21~m, reflecting the inherent spatial granularity of the database, while FM-based retrieval attains 1.36~m, raw CSI retrieval reaches 2.93~m with the Euclidean distance and 2.26~m with the cosine similarity, and random retrieval remains around 16.2~m. As $K$ grows, all curves rise as more distant samples are inevitably included. Nevertheless, FM-based retrieval consistently outperforms both raw CSI retrieval variants, which will translate into substantial performance gains for the subsequent ICL localization. In addition, a gap from FM-based retrieval to Perfect Retrieval still remains, since it is highly challenging for any representation-based strategy to match direct access to ground-truth UE positions.

\color{black}

\subsubsection{Comparison with Baselines}
\begin{figure}[t]
    \centering
    {\includegraphics[scale=0.8]{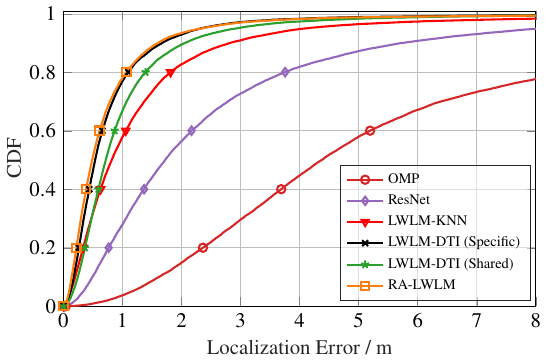}}
\caption{CDF of localization errors under the SS setting. }
\label{fig:SS_CDF}
\end{figure}

\begin{figure}[t]
    \centering
    {\includegraphics[scale=0.8]{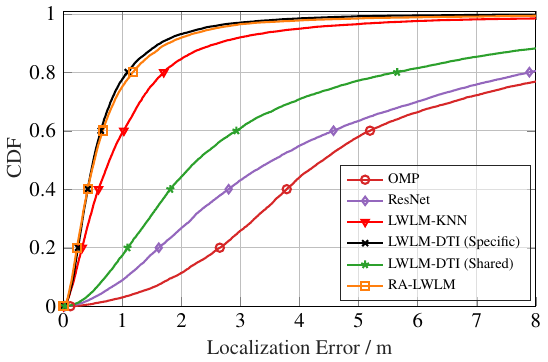}}
\caption{CDF of localization errors under the US setting. }
\label{fig:US_CDF}
\end{figure}

We compare the proposed RA-LWLM with the baseline methods. Fig.~\ref{fig:SS_CDF} and Fig.~\ref{fig:US_CDF} show the cumulative distribution function (CDF) of the localization error under the SS and US settings, respectively. We compare RA-LWLM against the following methods:
\begin{itemize}
    \item \textbf{OMP}: A model-based localization algorithm that uses orthogonal matching pursuit to estimate the angle of arrival and propagation delay of the dominant path between the UE and the BS, from which the UE position is computed via geometric triangulation. This method involves no training and operates purely on the query channel.
    \item \textbf{ResNet}: A supervised ResNet-34 backbone~\cite{wu2021learning} trained with BS configurations as additional input. This serves as a representative end-to-end learning baseline, jointly trained on the training samples from all SS and then directly tested on the US without retraining.
    
    \item \textbf{LWLM-DTI (shared)}: An LWLM encoder pretrained with the DTI objective and then fine-tuned end-to-end for localization, with the BS configuration $\bm{c}_s$ provided as additional input to the localization decoder~\cite{pan2025large}. The model is jointly fine-tuned on the training samples from all SS and then directly tested on both the SS and the US without any retraining. This reflects the performance achieved when an end-to-end fingerprint localization method is used to address the cross-scene localization problem.
    
    \item \textbf{LWLM-DTI (specific)}: The same LWLM-DTI architecture as LWLM-DTI (shared), fine-tuned end-to-end for localization with the BS configuration $\bm{c}_s$ provided as additional input to the localization decoder~\cite{pan2025large}. Different from LWLM-DTI (shared), an independent model is fine-tuned and evaluated within each individual scene, using only that scene's own training samples and requiring no cross-scene generalization in all scenes. This reflects the optimistic performance of FM-based end-to-end localization when a dedicated model is available per scene, at the cost of per-scene retraining.
    
    \item \textbf{LWLM-KNN}: We use the encoder of LWLM-DTI to extract channel representations and apply 1-nearest-neighbor retrieval over the per-scene database for query localization. This baseline isolates the contribution of in-context reasoning beyond pure retrieval on the same encoder backbone.
\end{itemize}

% \begin{figure}[t]
% \centering
% \includestandalone[scale=0.80]{Figures_Tikz/SS_CDF}
% \caption{CDF of localization errors under the SS setting. }
% \label{fig:SS_CDF}
% \end{figure}

% \begin{figure}[t]
% \centering
% \includestandalone[scale=0.80]{Figures_Tikz/US_CDF}
% \caption{CDF of localization errors under the US setting. }
% \label{fig:US_CDF}
% \end{figure}

As shown in Fig.~\ref{fig:SS_CDF}, under the SS setting, OMP, ResNet, and LWLM-DTI (shared) achieve median localization errors of 4.37~m, 1.72~m, and 0.72~m, respectively. The retrieval-based LWLM-KNN attains a median error of 0.82~m even with a simple KNN retrieval mechanism, incurring only a 13.9\% degradation relative to LWLM-DTI (shared). It is worth emphasizing that, once pretrained, LWLM-KNN requires no further parameter updates, which indicates that FM-based retrieval inherently possesses the potential for training-free cross-scene adaptation. Building on this, the proposed RA-LWLM further achieves a median localization error of 0.49~m, reducing the error by 88.8\%, 71.5\%, 31.9\%, and 40.2\% over OMP, ResNet, LWLM-DTI (shared), and LWLM-KNN, respectively. Furthermore, the proposed RA-LWLM achieves performance very close to LWLM-DTI (specific), indicating that under the SS setting, the retrieval-based approach incurs no performance loss compared with a dedicated per-scene model. It is worth noting that RA-LWLM does not require fine-tuning the FM when training the ICL module, whereas LWLM-DTI (specific) needs to fine-tune the FM itself during per-scene adaptation. This is an important advantage, since the FM representation is not distorted by the localization task and can be shared with other channel-related tasks. These results clearly demonstrate the performance advantage of RA-LWLM under the SS setting.

Under the US setting, as shown in Fig.~\ref{fig:US_CDF}, OMP can achieve a median localization error of 4.40~m, which is close to the results of the SS setting since it doesn't need training. The end-to-end trained ResNet and LWLM-DTI (shared) achieve median localization errors of 3.55~m and 2.27~m, respectively, corresponding to performance degradations of 106.4\% and 215.3\% compared to their SS counterparts. This indicates that end-to-end training fails to provide sufficient generalization capability when no retraining is performed on the new scene. In contrast, LWLM-KNN attains a median error of 0.78~m, which is nearly identical to its SS performance, while the proposed RA-LWLM achieves 0.53~m with only an 8.2\% degradation relative to the SS setting. Moreover, RA-LWLM outperforms ResNet and LWLM-DTI (shared) by 85.1\% and 76.7\% under the US setting, respectively, and achieves a median error almost identical to that of LWLM-DTI (specific). These results demonstrate the proposed RA-LWLM's superior cross-scene generalization capability.

% \begin{figure}[t]
% \centering
% \includestandalone[scale=0.80]{Figures_Tikz/training_scenes}
% \caption{Mean localization error of RA-LWLM versus the number of training scenes $|\mathcal{S}_{\text{train}}|$ on the SS and US test sets.}
% \label{fig:training_scenes_results}
% \end{figure}

\begin{figure}[t]
    \centering
    {\includegraphics[scale=0.8]{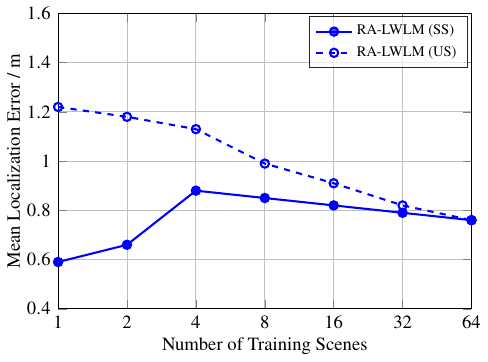}}
\caption{Mean localization error of RA-LWLM versus the number of training scenes $|\mathcal{S}_{\text{train}}|$ on the SS and US test sets.}
\label{fig:training_scenes_results}
\end{figure}

\subsubsection{Results with Different Numbers of Training Scenes}

We vary $|\mathcal{S}_{\text{train}}| \in \{1,\,2,\,4,\,8,\,16,\,32,\,64\}$ with $N^{\mathcal{D}}_s = 4000$ and all other hyperparameters fixed, and show the mean localization error on both SS and US sets in Fig.~\ref{fig:training_scenes_results}. The two curves exhibit distinct trends. The US error decreases monotonically as the number of training scenes increases, dropping from 1.22~m at $|\mathcal{S}_{\text{train}}| = 1$ to 0.76~m at $|\mathcal{S}_{\text{train}}| = 64$. The SS error, in contrast, first rises from 0.59~m to 0.88~m as $|\mathcal{S}_{\text{train}}|$ grows from 1 to 4, and then decreases monotonically to 0.76~m at $|\mathcal{S}_{\text{train}}| = 64$. The initial SS dip reflects in-scene overfitting when very few training scenes are available. In this regime, the ICL module memorizes the geometry and propagation patterns of these specific scenes, achieving low in-scene error but failing to generalize to unseen scenes, as evidenced by the US error remaining above 1.13~m for $|\mathcal{S}_{\text{train}}| \le 4$. As more training scenes are added, the ICL module is forced to learn scene-agnostic spatial reasoning patterns rather than memorize scene-specific structures, which gradually reduces the US error. Furthermore, the gap between SS and US shrinks rapidly with $|\mathcal{S}_{\text{train}}|$. With one training scene, the SS-US gap reaches 106.8\%, with the US error more than twice the SS error. The gap shrinks to 11.0\% at 16 scenes, 3.8\% at 32 scenes, and is fully closed at 64 scenes, where the two curves converge to the same level of 0.76~m. This convergence confirms that RA-LWLM has effectively learned to generalize across scenes rather than memorize them, and that in our setting, training with around 32 to 64 scenes is sufficient to achieve nearly lossless cross-scene generalization.

% \begin{figure}[t]
% \centering
% \includestandalone[scale=0.80]{Figures_Tikz/training_dataset}
% \caption{Mean localization error of RA-LWLM versus the per-scene database size $N^{\text{data}}_s$ on the SS and US test sets.}
% \label{fig:training_dataset_results}
% \end{figure}

\begin{figure}[t]
    \centering
    {\includegraphics[scale=0.8]{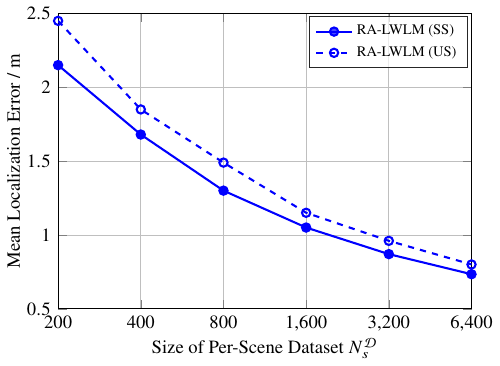}}
\caption{Mean localization error of RA-LWLM versus the per-scene database size $N^{\text{data}}_s$ on the SS and US test sets.}
\label{fig:training_dataset_results}
\end{figure}

\subsubsection{Results with Different Sizes of Reference Datasets}

We fix the number of training scenes to 16 and vary the per-scene database size $N^{\mathcal{D}}_s \in \{200,\,400,\,800,\,1600,\,3200,\,6400\}$, evaluating RA-LWLM on both SS and US test sets. Fig.~\ref{fig:training_dataset_results} reports the mean localization error as a function of $N^{\mathcal{D}}_s$. As shown in the figure, the localization error on both SS and US decreases monotonically as the per-scene database grows. Specifically, the SS error drops from 2.15~m at $N^{\mathcal{D}}_s = 200$ to 0.73~m at $N^{\mathcal{D}}_s = 6400$, corresponding to a 66.0\% reduction. The US error follows almost the same trend, decreasing from 2.45~m to 0.80~m, with a 67.3\% reduction. This trend is expected, as a denser database yields more spatially faithful references and thus more informative in-context examples for the ICL module. It is also worth noting that the SS-US gap remains small and stable across the entire range of $N^{\mathcal{D}}_s$, fluctuating only between 9\% and 15\%. This is in contrast to the behavior observed when increasing the number of training scenes. The two experiments together reveal a clear separation of roles: the number of training scenes governs how well the ICL module generalizes to unseen environments, whereas the per-scene database size governs the absolute localization accuracy attainable in any given scene. From a deployment perspective, the operator can therefore flexibly trade off labeling cost against localization accuracy by collecting more fingerprints in any target scene, without retraining the FM encoder, the ICL module, or the routing module.

% \begin{figure}[t]
% \centering
% \includestandalone[scale=0.80]{Figures_Tikz/training_data}
% \caption{Mean localization error of RA-LWLM versus the number of training scenes under a fixed total labeling budget of 80{,}000 labels.}
% \label{fig:scene_density_tradeoff}
% \end{figure}

\begin{figure}[t]
    \centering
    {\includegraphics[scale=0.8]{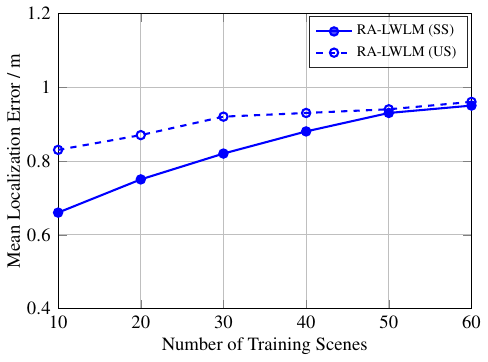}}
\caption{Mean localization error of RA-LWLM versus the number of training scenes under a fixed total labeling budget of 80{,}000 labels.}
\label{fig:scene_density_tradeoff}
\end{figure}

\begin{table*}[t]
\centering
\caption{Localization error of the RA-LWLM and the baselines under different BS configurations}
\label{tab:bs_config}
\begin{tabular}{cc|ccc|ccc|ccc|ccc|ccc}
\toprule
\multirow{2}{*}{$N^{\text{ant}}_s$} & \multirow{2}{*}{$B^{\text{bw}}_s$} & \multicolumn{3}{c|}{OMP} & \multicolumn{3}{c|}{ResNet} & \multicolumn{3}{c|}{LWLM-DTI (shared)} & \multicolumn{3}{c|}{LWLM-KNN} & \multicolumn{3}{c}{\textbf{RA-LWLM}} \\
\cmidrule(lr){3-5} \cmidrule(lr){6-8} \cmidrule(lr){9-11} \cmidrule(lr){12-14} \cmidrule(lr){15-17}
 & & Mean & Med. & p90 & Mean & Med. & p90 & Mean & Med. & p90 & Mean & Med. & p90 & Mean & Med. & p90 \\
\midrule
\multirow{3}{*}{$8$}  & $5$  & 8.25 & 7.20 & 14.56 & 5.66 & 4.85 & 11.28  & 4.13 & 3.29    & 8.07 & 1.77  & 1.20 & 3.49 & \textbf{1.21} & \textbf{0.77} & \textbf{2.40} \\
                      & $10$ & 5.82 & 4.52 &  11.18 & 4.77  &  3.40 &  10.39 & 3.75 &  2.59 & 8.29 & 1.52 &  0.95 & 3.00 & \textbf{1.17} & \textbf{0.71} & \textbf{2.37} \\
                      & $20$ & 4.87 & 3.19 &   10.91 & 6.79 & 5.63 & 12.94  & 3.93 & 2.65 & 8.30 & 1.21 &  0.75 & 2.44 & \textbf{0.97} & \textbf{0.57} & \textbf{2.03} \\
\midrule
\multirow{3}{*}{$16$} & $5$  & 8.32 & 7.19 & 15.55 & 6.27 & 4.73 & 13.88 & 4.60 & 2.51 & 12.02 & 1.86 & 1.00  & 4.03 & \textbf{1.30} & \textbf{0.78} & \textbf{2.57} \\
                      & $10$ & 5.45 & 4.04 & 10.68 & 4.76 & 3.81 & 9.76 & 3.45 & 2.27 & 7.77 & 1.13 &  0.72 & 2.20 & \textbf{0.74} & \textbf{0.45} & \textbf{1.45}  \\
                      & $20$ & 4.63 & 3.07 &  10.12 & 5.35 & 3.90 & 12.08 & 4.24 & 2.59 & 10.56 & 1.00 & 0.64 & 1.97 & \textbf{0.77} & \textbf{0.47} & \textbf{1.64} \\
\midrule
\multirow{3}{*}{$32$} & $5$  & 7.94 & 6.63 & 14.17 & 8.36 & 7.97 & 14.45 & 5.92 & 5.40 & 10.33 & 1.92 & 0.96 & 4.35 & \textbf{1.14} & \textbf{0.70} & \textbf{2.26} \\
                      & $10$ & 4.80 & 3.88 & 8.36 & 5.07 & 3.50 & 11.64 & 4.22 & 2.77 & 9.98 & 1.24 & 0.74 & 2.62 & \textbf{0.74} & \textbf{0.45} & \textbf{1.46} \\
                      & $20$ & 4.44 & 2.92 &  11.10 &  5.60 & 3.92 & 12.90 & 4.41 & 2.71 & 11.39 & 0.90 & 0.59 & 1.88 & \textbf{0.61} & \textbf{0.41} & \textbf{1.27}\\
\midrule
\multicolumn{2}{c|}{Average} & 6.06 & 4.74 & 11.85 & 5.85&  4.63 & 12.15 & 4.29 & 2.98 & 9.63 & 1.39  & 0.84 & 2.89 & \textbf{0.96} & \textbf{0.59} & \textbf{1.94} \\
\bottomrule
\end{tabular}
\end{table*}

\subsubsection{Trade-off between Scene Diversity and Per-Scene Density}

The two previous experiments varied $|\mathcal{S}_{\text{train}}|$ and $N^{\mathcal{D}}_s$ independently. When the total labeling budget $N_{\text{total}} = |\mathcal{S}_{\text{train}}| \cdot N^{\mathcal{D}}_s$ is fixed, the operator faces a trade-off between more scenes with lower per-scene density and fewer scenes with higher density. To examine this, we fix $N_{\text{total}} = 80{,}000$ and vary $|\mathcal{S}_{\text{train}}| \in \{10,\,20,\,30,\,40,\,50,\,60\}$, scaling $N^{\mathcal{D}}_s$ inversely. As shown in Fig.~\ref{fig:scene_density_tradeoff}, the SS error grows monotonically from 0.66~m to 0.95~m, since sparser per-scene sampling degrades the spatial fidelity of the retrieved references. The US error also grows from 0.83~m to 0.96~m and is minimized at $|\mathcal{S}_{\text{train}}| = 10$, indicating that the retrieval-quality loss from reduced density outweighs the gain from added diversity. More importantly, the SS-US gap narrows from 25.8\% at 10 scenes to 1.1\% at 50 scenes, where both curves converge around 0.95~m. The results also show that scene diversity governs cross-scene transferability while per-scene density governs absolute accuracy, and the two effects compete under a fixed budget.

\subsubsection{Detailed Comparison across BS Configurations}
This experiment provides a per-configuration comparison of RA-LWLM and the baselines under different BS configurations. All models are trained on 20 scenes with 4{,}000 samples per scene and evaluated on the US test set. For each BS configuration $(N^{\text{ant}}_s, B^{\text{bw}}_s)$, we randomly generate 5 scenes with different building layouts and BS heights, and record the averaged results. As shown in Table~\ref{tab:bs_config}, overall, the results of the proposed RA-LWLM follow the expected trend that larger bandwidth and more antennas yield higher localization accuracy, since higher angular and delay resolution produce more discriminative channel representations. However, there are some exceptions. For example, the proposed RA-LWLM achieves a lower mean error with 16 antennas and 10~MHz bandwidth than with 16 antennas and 20~MHz bandwidth. This is because the building layouts and the fingerprint sampling are randomized per configuration, and when some scenes contain richer multipath or larger NLOS regions, the gain from a better BS configuration cannot fully compensate for the inherent localization difficulty of those scenes. Such fluctuations are more pronounced for the poorly-generalizing ResNet and LWLM-DTI (shared) baselines. Despite these per-configuration variations, the comparison remains meaningful. We can find that RA-LWLM achieves the lowest mean error in every configuration, with an overall average of 0.96~m, compared with 1.39~m for LWLM-KNN, 4.29~m for LWLM-DTI (shared), 5.85~m for ResNet, and 6.06~m for OMP, confirming the robustness of the retrieval-augmented design across a wide range of hardware budgets and scene complexities.

\subsubsection{Ablation Study}
To isolate the contribution of each component, we evaluate four ablated variants of RA-LWLM and two additional routing configurations. The mean, median, and 90\% localization errors on both SS and US test sets are reported in Table~\ref{tab:ablation}. The ablated variants are (i) \textbf{w/o FM}, where the pretrained LWLM encoder is replaced with raw CSI retrieval, removing the contribution of self-supervised representation learning; (ii) \textbf{w/o MoE}, where the multi-expert design is replaced by a single expert that consumes all retrieved references, removing the per-query context-size specialization; (iii) \textbf{w/o selector}, where the learned router is replaced by uniform routing with average score; and (iv) \textbf{w/o centering}, where the ICL transformer directly predicts the absolute UE position rather than a centered-normalized residual on top of the weighted centroid. We additionally evaluate two hard-routing variants, i.e., RA-LWLM top-1 and RA-LWLM top-2, which restrict the selector to the single most-weighted expert and the two most-weighted experts, respectively, instead of the soft combination over all experts. As shown in Table~\ref{tab:ablation}, every component contributes a measurable gain, but their roles are clearly distinct. Removing the FM encoder causes the most severe degradation, increasing the mean error from 0.75~m to 1.01~m on SS and from 0.87~m to 1.33~m on US, corresponding to relative increases of 34.7\% and 52.9\%, respectively. This confirms that the pretrained channel representation is the foundation of the entire pipeline, since without a transferable feature space, neither retrieval nor in-context reasoning can operate effectively. The MoE design is the second most impactful component, and removing it raises the mean error by 25.3\% on SS and 18.4\% on US, validating that adapting the context size to each query is essential for handling the heterogeneous propagation conditions and varying retrieval quality across queries. Replacing the learned selector with uniform routing degrades performance by 18.7\% on SS and 11.5\% on US, indicating that learning query-aware routing weights over the experts cannot be replaced by a static uniform combination. Removing position centering causes a 10.7\% degradation on SS and a 10.3\% degradation on US, since predicting absolute coordinates causes the ICL transformer to overfit to the coordinate range of the training scenes, whereas centered-normalized residual prediction forces it to learn transferable spatial patterns among the retrieved references. Finally, the two hard-routing variants confirm the value of soft combination. RA-LWLM top-1 incurs a 6.7\% SS and 8.0\% US gap, while RA-LWLM top-2 shrinks this gap to 2.7\% and 4.6\%. The full soft combination is best, showing that several experts at different context sizes provide complementary information that a single dominant expert cannot capture.

\color{black}

\begin{table}[t]
\centering
\caption{Ablation study of RA-LWLM on the SS and US test sets.}
\label{tab:ablation}
\begin{tabular}{l|ccc|ccc}
\toprule
\multirow{2}{*}{\textbf{Variant}} & \multicolumn{3}{c|}{\textbf{SS}} & \multicolumn{3}{c}{\textbf{US}} \\
 & Mean & Med. & p90 & Mean & Med. & p90 \\
\midrule
w/o FM  & 1.01 & 0.70 & 2.01  & 1.33 & 0.81 & 2.85 \\
w/o MoE & 0.94 & 0.67 & 1.92 & 1.03 & 0.70 & 2.09  \\
w/o selector & 0.89 & 0.64 & 1.93 & 0.97 & 0.68 & 1.99  \\
w/o centering & 0.83 & 0.53 & 1.60 & 0.96 & 0.62 & 1.93 \\
RA-LWLM top-1 & 0.80 & 0.49 & 1.58 & 0.94 & 0.57 & 1.93  \\
RA-LWLM top-2 & 0.77 & 0.47 & 1.54 & 0.91 & 0.55 & 1.86  \\
\textbf{RA-LWLM}              & \textbf{0.75} & \textbf{0.45} & \textbf{1.52} & \textbf{0.87} & \textbf{0.53} & \textbf{1.82} \\
\bottomrule
\end{tabular}
\end{table}

\section{Conclusion and Future Directions}
\label{sec:Conclusion}

In this paper, we proposed RA-LWLM, a retrieval-augmented in-context localization framework that decouples scene-invariant channel representation learning from scene-specific position inference, enabling training-free adaptation to new scenes by simply refreshing a per-scene fingerprint database. Beyond confirming the effectiveness of this design on heterogeneous ray-tracing benchmarks, our experiments uncovered an asymmetry between the two axes of training resources, where scene diversity governs cross-scene transferability while per-scene database density governs absolute in-scene accuracy. This separation translates into a concrete data collection guideline, where moderate scene diversity combined with dense per-scene sampling yields a balanced operating point well-suited for scalable 6G localization deployments.

Several directions remain open for future work. First, the FM encoder is currently pretrained with a single self-supervised objective, and identifying the optimal pretraining objective tailored to retrieval-based localization is an interesting direction. Second, the per-scene database is currently constructed by uniform sampling, and the relationship between database construction and the spatial distribution of references deserves a dedicated study, since adaptive sampling could deliver the same retrieval quality with substantially fewer labeled samples. Third, the database is assumed to be static in this work, while in practice, the propagation environment evolves over time as buildings, foliage, and street layouts change. Developing efficient mechanisms to update the database in response to such environmental drift is also an important direction. Fourth, the retrieval module currently relies on a fixed Euclidean similarity in the representation space, and designing more efficient and query-adaptive retrieval strategies could further improve both the quality of the in-context references and the scalability to large per-scene databases. Finally, extending the framework to multi-BS localization, mobile UEs with temporal context, and real-world measurement campaigns would further validate the practicality of the proposed retrieval-augmented in-context paradigm in 6G networks.

% \ifCLASSOPTIONcaptionsoff
% \newpage
% \fi
\balance 
% \end{thebibliography}
\bibliographystyle{IEEEtran}
\bibliography{IEEEabrv, ref}

\end{document}